\documentclass [preprint,prb,aps,letterpaper,showpacs,showkeys,onecolumn,amsmath,amssymb,floatfix] {revtex4}
\usepackage{graphicx}
\usepackage{color}
\usepackage{textcomp}
\usepackage{mathptmx} 

\begin{document}
\vskip 0.4cm

\title{Electron transport in a ferromagnetic/normal/ferromagnetic tunnel junction based on the
surface of a topological insulator}

\author{ Jian-Hui Yuan}
\altaffiliation{The corresponding author, E-mail:
jianhui831110@163.com}
\author{Yan Zhang}

\affiliation{The department of Physics, Guangxi medical university,
Nanning, Guangxi, 530021, China}
\author{ Jian-Jun Zhang}
\author{ Ze Cheng}

\affiliation{ School of Physics, Huazhong University of Science and
Technology,  Wuhan 430074, China}
\date{\today}

\begin{abstract}{ {We theoretically study the electron transport
properties in a ferromagnetic/normal/ferromagnetic tunnel junction,
which is deposited on the top of a topological surface. The
conductance at the parallel (\textbf{P}) configuration can be much
bigger than that at the antiparallel (\textbf{AP}) configuration.
Compared \textbf{P } with \textbf{ AP} configuration,  there exists
a shift of phase which can be tuned by gate voltage. We find that
the exchange field weakly affects the conductance of carriers for
\textbf{P} configuration but can dramatically suppress the
conductance of carriers for \textbf{AP} configuration. This
controllable electron transport  implies anomalous magnetoresistance
in this topological spin valve, which may contribute to the
development of spintronics . In addition, we find that there is a
Fabry-Perot-like electron interference.
 }}
\end{abstract}

\maketitle
\section{\textbf{Introduction}}
The concept of a topological insulator (\textbf{TI}) dates back to
the work of Kane and Mele, who focused on two-dimensional
(\textbf{2D}) systems $^{1}$. There has been much recent interest in
\textbf{TIs}, three-dimensional insulators with metallic surface
states protected by time reversal invariance ${[1-25]}$. Its
theoretical ${[2]}$ and experimental ${[3]}$ discovery has
accordingly generated a great deal of excitement in the condensed
matter physics community. In particular, the surface of a
three-dimensional (\textbf{3D}) \textbf{TI}, such as
Bi$_{2}$Se$_{3}$ or Bi$_{2}$Te$_{3}$ ${[4]}$, is a 2D metal, whose
band structure consists of an odd number of Dirac cones, centered at
time reversal invariant momenta in the surface Brillouin zone
${[5]}$. This corresponds to the infinite mass Rashba model ${[6]}$,
where only one of the spin-split bands exists. This has been
beautifully demonstrated by the spin- and angle-resolved
photoemission spectroscopy ${[7,8]}$.  Surface sensitive experiments
such as angle-resolved photoemission spectroscopy (\textbf{ARPES})
and scanning tunneling microscopy (\textbf{STM}) ${[9,10]}$ have
confirmed the existence of this exotic surface metal, in its
simplest form, which takes a single Dirac dispersion. Recent
theoretical and experimental discovery of the two dimensional
(\textbf{2D}) quantum spin Hall system ${[11-18]}$ and its
generalization to the \textbf{TI} in three dimensions ${[19-21]}$
have established the state of matter in the time-reversal symmetric
systems.

The time-reversal invariant \textbf{TI} is a new state of matter,
distinguished from a regular band insulator by a nontrivial
topological invariant, which characterizes its band structure
${[11]}$. Currently, most works focus on searching for TI materials
and novel transport properties. To my knowledge, the fabrication of
such TI-based nanostructure is still a challenging task. Usually
such structures are fabricated by utilizing the split gate and
etching technique ${[22]}$. On the other hand,  the \textbf{3D}
\textbf{TIs} are expected to show several unique properties when the
time reversal symmetry is broken ${[23-25]}$. This can be realized
directly by a ferromagnetic insulating (\textbf{FI}) layer attached
to the \textbf{3D} \textbf{TI} surface.  One remarkable feature of
the Dirac fermions is that the Zeeman field acts like a vector
potential: the Dirac Hamiltonian is transformed as $\sigma \cdot
\textbf{k}\longrightarrow\sigma \cdot(\textbf{k+H})$ by the Zeeman
field $\textbf{H}$ ${[26]}$. This is in contrast to the
Schr$\ddot{o}$dinger electrons in conventional semiconductor
heterostructures modulated by nanomagnets ${[27-29]}$.

 In this work, we  study the electron transport
properties in a ferromagnetic (\textbf{F})/normal/\textbf{N})/
ferromagnetic (\textbf{F}) tunnel junction, which is deposited on
the top of a topological surface.  Ferromagnetic Permalloy
electrodes are formed by electron-beam lithography (EBL) followed by
thermal evaporation; a second EBL step establishes contact to the
Permalloy via Cr /Au electrodes ${[30]}$. As shown in Fig.1, the
\textbf{FI} is put on the top of the \textbf{TI} to induce an
exchange field via the magnetic proximity effect. The easy axis of a
\textbf{FI} stripe is usually along its length direction and thus
either in parallel (\textbf{P}) or antiparallel (\textbf{AP}) with
the $+y$ axis. We find that the conductance at the  \textbf{P}
configuration can be much bigger than that at the  \textbf{AP}
configuration. Compared \textbf{P } with \textbf{ AP} configuration,
there exists a shift of phase which can be tuned by gate voltage. We
find that the exchange field weakly affects the conductance of
carriers for \textbf{P} configuration but can dramatically suppress
the conductance of carriers for \textbf{AP} configuration. This
controllable electron transport implies anomalous magnetoresistance
in this topological spin valve, which may contribute to the
development of spintronics. Compared with the conventional
\textbf{F/N/F} tunneling based on two dimensional electron gas
(\textbf{2DEG}), the result implies  the existence of
Fabry-Perot-like electron interference in \textbf{F/N/F} based on
the \textbf{TI}. In Sec. \uppercase\expandafter{\romannumeral 2} ,
we introduce the model and method for our calculation. In Sec.
\uppercase\expandafter{\romannumeral 3}, the numerical analysis to
our important  issues is reported. Finally, a brief summary is given
in sec. \uppercase\expandafter{\romannumeral 4}.

\section{ model and method}
\begin{figure}
\resizebox{1.0\columnwidth}{!}{%
  \includegraphics{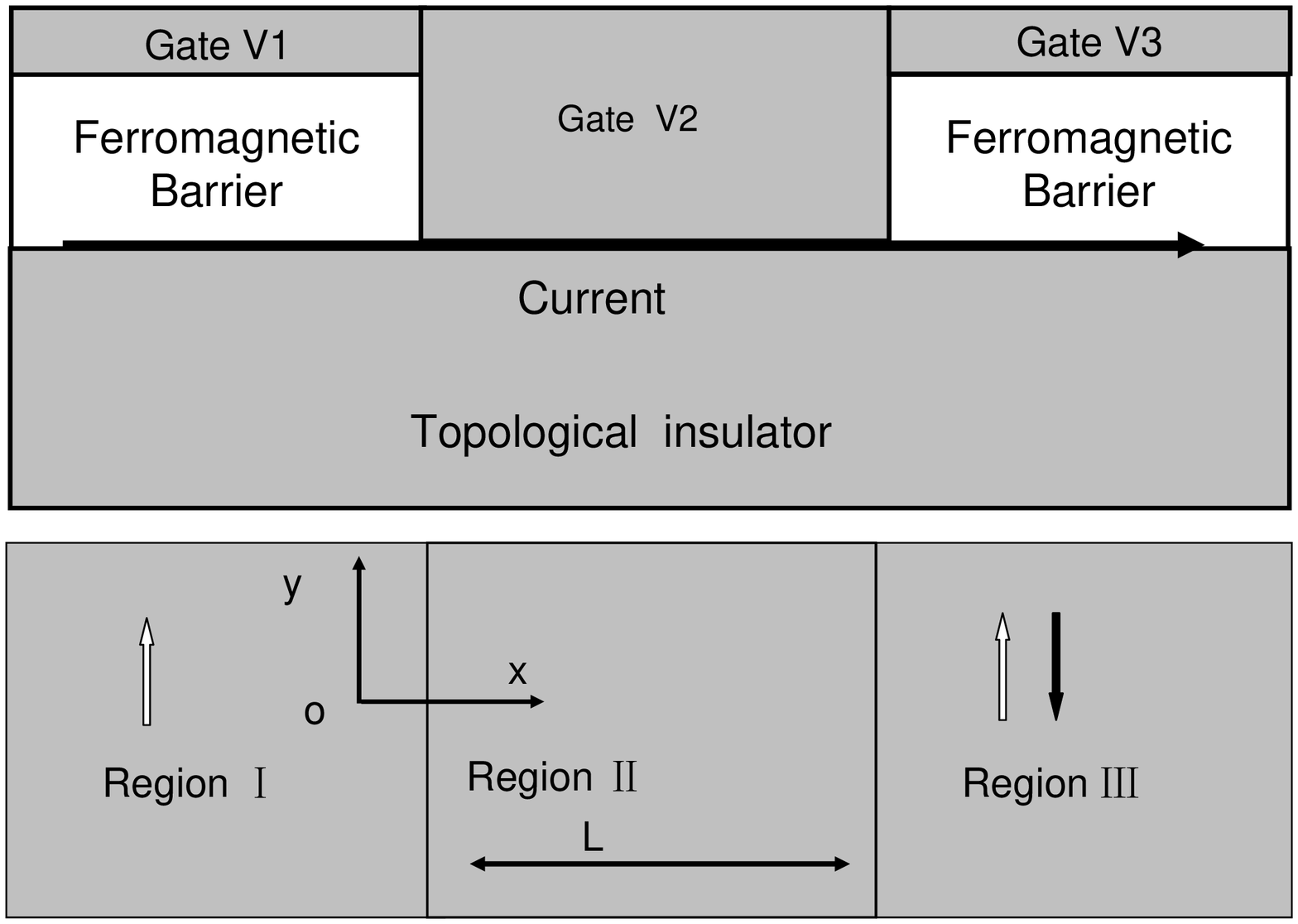}
}
\caption{ Schematic illustration of the device. Top: Schematic
diagram of two ferromagnetic barriers on the topological surface
divided by  a gate electrode  at a distance $L$.   Bottom: The
magnetization directions of adjacent \textbf{FI} stripes are
parallel (\textbf{P}) in the configuration  and antiparallel
(\textbf{AP}) in the configuration.  }\label{FIG:2eps}       
\end{figure}

Now, let us consider a \textbf{F}/\textbf{N}/ \textbf{F} tunnel
junction which is deposited on the top of a topological surface
where a gate electrode is attached to the ferromagnetic material.
The ferromagnetism is induced due to the proximity effect by the
ferromagnetic insulators deposited on the top as shown in Fig. 1. We
assume that the initial magnetization of \textbf{ FI} stripes in the
region \uppercase\expandafter{\romannumeral 1} is aligned with the
+y axis. In an actual experiment, one can use a magnet with very
strong (soft) easy axis anisotropy to control the ferromagnetic
material. Thus we focus on charge transport at the Fermi level of
the surface of \textbf{TIs}, which is described by the 2D Dirac
Hamiltonian
\begin{eqnarray}
H = \upsilon_{F} \sigma \cdot \textbf{p} + \sigma \cdot \textbf{M} +
V{(x)},
\end{eqnarray}\label{1}
where  $\sigma$ is Pauli matrices , $\textbf{M} = M_y(x)= M_0
(\Theta(-x)+ \gamma \Theta(x-L))$ is the effective exchange field
and $V(x)=U_g \Theta(x)\Theta(L-x)+ V_g \Theta(x-L)$ is the gate
voltage, where $\gamma =+1$ ($-1$) corresponds to the \textbf{P}
(\textbf{AP}) configurations of magnetization  and $\Theta(x)$ is
the Heaviside step function.

Because of the translational invariance of the system along $y$
direction, the equation $H \Psi(x, y)=E \Psi(x, y)$ admits solutions
of the form $\Psi(x,y)=(\Psi_{1}(x), \Psi_{2}(x))^{T}\exp(i k_y y)$.
We set $\hbar = \upsilon_F =1$ in the following. Then, with the
above Hamiltonian, the wave function in the whole system is given by
\begin{eqnarray}
\Psi_{1}&=&\left\{\begin{array}{lllll  }  \exp (i k_{x_1} x) + r
\exp (-i k_{x_1} x), &
\quad x<0, \\
\\
a \exp (i q_{x} x) + b \exp (-i q_{x} x), & \quad 0<x<L,
\\
\\
t \exp (i k_{x_2} (x-L)) , & \quad x>L,\end{array} \right.
\end{eqnarray}\label{2}

\begin{eqnarray}
\Psi_{2}&=&\left\{\begin{array}{lllll  } \alpha^{+} \exp (i k_{x_1}
x) + r \alpha^{-} \exp (-i k_{x_1} x), &
\quad x<0, \\
\\
a \beta^{+} \exp (i q_{x} x) + b \beta^{-} \exp (-i q_{x} x), &
\quad 0<x<L,
\\
\\
t \alpha \exp (i k_{x_2} (x-L)) , & \quad x>L,\end{array} \right.
\end{eqnarray}\label{2}
where $k_{x_1}=E \cos \theta_{F_1}$, $q_{x} =  (E-U_{g})\cos \theta$
and $k_{x_2}=(E-V_{g})\cos \theta_{F_2}$ are wave vectors in region
\uppercase\expandafter{\romannumeral 1}, region
\uppercase\expandafter{\romannumeral 2} and region
\uppercase\expandafter{\romannumeral 3}, $\alpha^{\pm}=\pm \exp (\pm
i \theta_{F_1})$, $\beta^{\pm}=\pm \exp (\pm i \theta)$ and $\alpha=
\exp (i \theta_{F_2})$. The momentum $k_y$ conservation should be
satisfied everywhere such as $k_y = E \sin \theta_{F_1}-M_0 =
(E-U_g)\sin \theta= (E-V_g) \sin \theta_{F_2}-\gamma M_0$. Also, $r$
and $t$ are reflection and transmission coefficients, respectively.
Continuities of the wave function $\Psi$ at $x=0$ and $x=L$ are
$\Psi(0^{-}) = \Psi(0^{+})$ and $ \Psi(L^{-}) = \Psi(L^{+})$,
respectively. We find that the transmitted electron coefficient
$t_\gamma$ is given by
\begin{eqnarray}
t_\gamma =\frac{2 \cos \theta_{F_1} \cos \theta \exp(-i
k_{x_2}L)}{s_{1, \gamma} \cos (q_x L) + i s_{2, \gamma} \sin (q_x
L)},
\end{eqnarray}\label{1}
with $s_{1, \gamma}= \cos \theta(\exp (i \theta_{F_2})+\exp(-i
\theta_{F_1}))$ and $s_{2, \gamma}= i \sin \theta (\exp (-i
\theta_{F_1})-\exp(i \theta_{F_2}))-\exp(i (\theta_{F_2} -
\theta_{F_1}))-1$. Then
\begin{eqnarray}
T_\gamma = |t_\gamma|^2 \Re(\cos \theta_{F_2}/\cos \theta_{F_1}),
\end{eqnarray}\label{1}
where the factor $\Re(\cos \theta_{F_2}/\cos \theta_{F_1})$ is due
to current conservation. In the linear transport regime and for low
temperature, we can obtain the conductance $G$ by introducing it as
the electron flow averaged over half the Fermi surface from the
well-known Landauer-Buttiker formula ${[25, 31, 32]}$
\begin{eqnarray}
G_\gamma \sim 1/2\int_{-\pi/2}^{\pi/2} T_\gamma (E_{F},E_{F}
\cos\theta_{F_1} ) \cos\theta_{F_1} d\theta_{F_1}.
\end{eqnarray}\label{5}

\section{Results and Discussions}

\begin{figure}
\resizebox{1.0\columnwidth}{!}{%
  \includegraphics{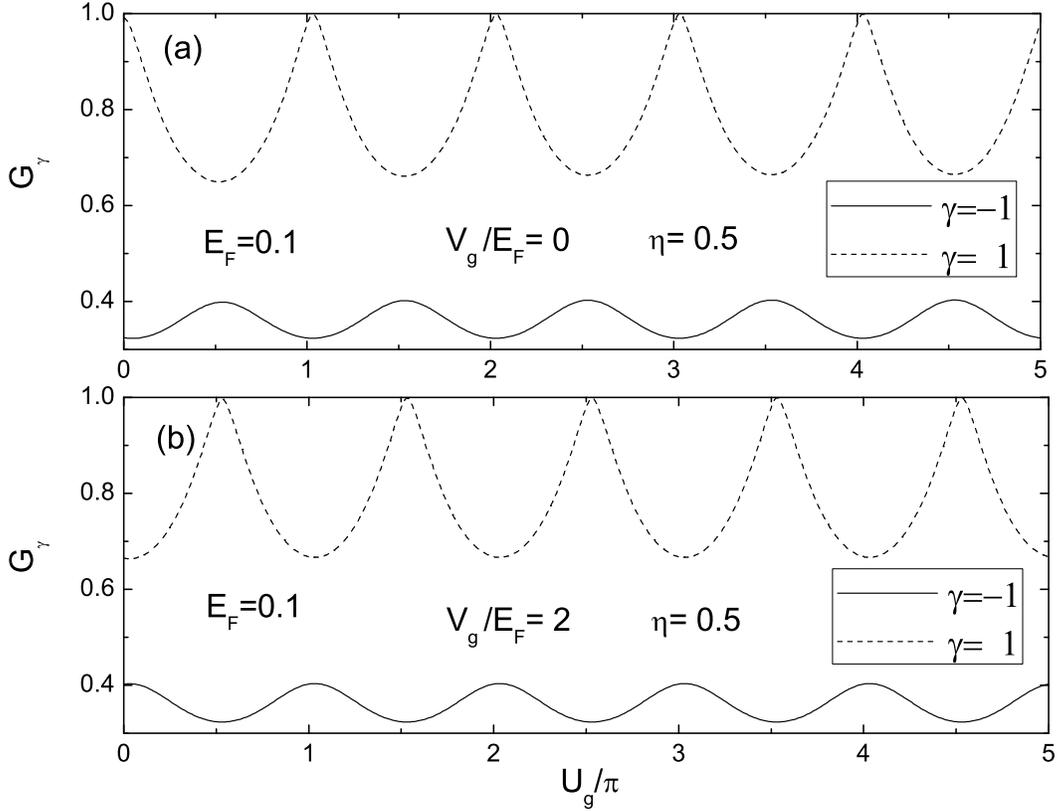}
}
\caption{ Gate voltage dependence of the conductances with a
\textbf{ P} ($\gamma=1$) and  \textbf{AP} ($\gamma=-1$)
configuration in the two cases: (a) $V_g /E_F =0$ and (b) $V_g /E_F
=2$. The values of the other parameters are $E_F =0.1$ and
$\eta=0.5$ }\label{FIG:2eps}       
\end{figure}

For convenience we express all quantities in dimensionless units by
means of the length of the basic unit $L$ and the energy $E_0=\hbar
v_F /L$. For a typical value of $L=50$ nm and the Bi$_2$Se$_3$
material $v_F=5\times10^5$ m/s, one has $E_0=6.6$ meV. We set the
energy of electron $E=E_F$ and also define the value $\eta$ with the
form $\eta=M_0 /E_F$ in our calculation.

In Fig.2, we show gate voltage dependence of the conductances with a
\textbf{ P} ($\gamma=1$) and  \textbf{AP} ($\gamma=-1$)
configuration in the two cases: (a) $V_g /E_F =0$ and (b) $V_g /E_F
=2$. The value of the other parameter is $E_F =0.1$ and $\eta=0.5$.
The presence of quantum modulation are seen in these two figures. We
can see an oscillation of the electrical conductance with a period
of $\pi$ when the voltage $U_g$ is larger than $E_F$. The
conductance at the \textbf{P} configuration can be much bigger than
that at the \textbf{AP} configuration. We find that a minimum of
conductance at the \textbf{P} configuration corresponds to a maximum
of conductance at the \textbf{AP} configuration [see in fig.2 (a)]
when the voltage $U_g$ is larger than $E_F$. In Fig. 2(b), a similar
tendency to Fig. 2(a) is seen. In distinct contrast to Fig.2(a), a
minimum of conductance at the \textbf{P} configuration here
corresponds to a maximum of conductance at the \textbf{AP}
configuration [see in fig.2 (b)]. That is to say, there exists a
shift of $\pi$-phase. To understand these results intuitively, we
consider that the gate voltage $U_g$ is larger than the Fermi energy
$E_F$. For  the given Fermi energy $E_F =0.1$, the condition $U_g
\gg E_F$ is easily satisfied. In this limit we have
$\theta\rightarrow 0$ and hence the transmission probability
$T_\gamma \sim (2 \cos^2 \theta_{F_1}/(1+ \cos \theta_{F_1}\cos
\theta_{F_2} -\cos (2 U_g L) \sin \theta_{F_1}\sin \theta_{F_2})
\Re(\cos \theta_{F_2}/\cos \theta_{F_1}) $. For $\gamma=1$ and $V_g
/ E_F = 0$ (or $2$), we find the $\theta_{F_1}\equiv \theta_{F_2}$
(or $-\theta_{F_2}$), and thus $T_\gamma \sim  \cos^2
\theta_{F_1}/(1- \cos^2 ( U_g L+ \delta) \sin^2 \theta_{F_1})$ where
$\delta =0$ (or $\pi/2$) corresponds to $V_g / E_F = 0$ (or $2$).
Thus the phase difference between $V_g / E_F =0$ and $V_g / E_F =2$
is given by $ U_g L$. We find $G_\gamma \propto \cos^2 (U_g L)$ for
$V_g / E_F = 0$ but $G_\gamma \propto \sin^2 ( U_g L)$ for  $V_g /
E_F = 2$. When $U_g L$ is equal to the half period of $\pi$,  a
minimum of conductance will appear for $V_g / E_F = 0$ but a maximum
of conductance will appear for $V_g / E_F = 2$. When $U_g L$ is
equal to the period of $\pi$,  a maximum of conductance will appear
for $V_g / E_F = 0$ but a minimum of conductance will appear for
$V_g / E_F = 2$. Furthermore, we find that $G_\gamma$ oscillates
between $2/3$ and $1$ for $\gamma=1$. For $\gamma=-1$ and $V_g / E_F
= 0$ (or $2$), there is a similar tendency to the case of
$\gamma=1$. We can see that $G_\gamma$ is suppressed obviously by
the strength of the effective exchange field . Nevertheless, there
exists a shift of $\pi$-phase because of the factor $\cos (2 U_g
L)$.
\begin{figure}
\resizebox{1.0\columnwidth}{!}{%
  \includegraphics{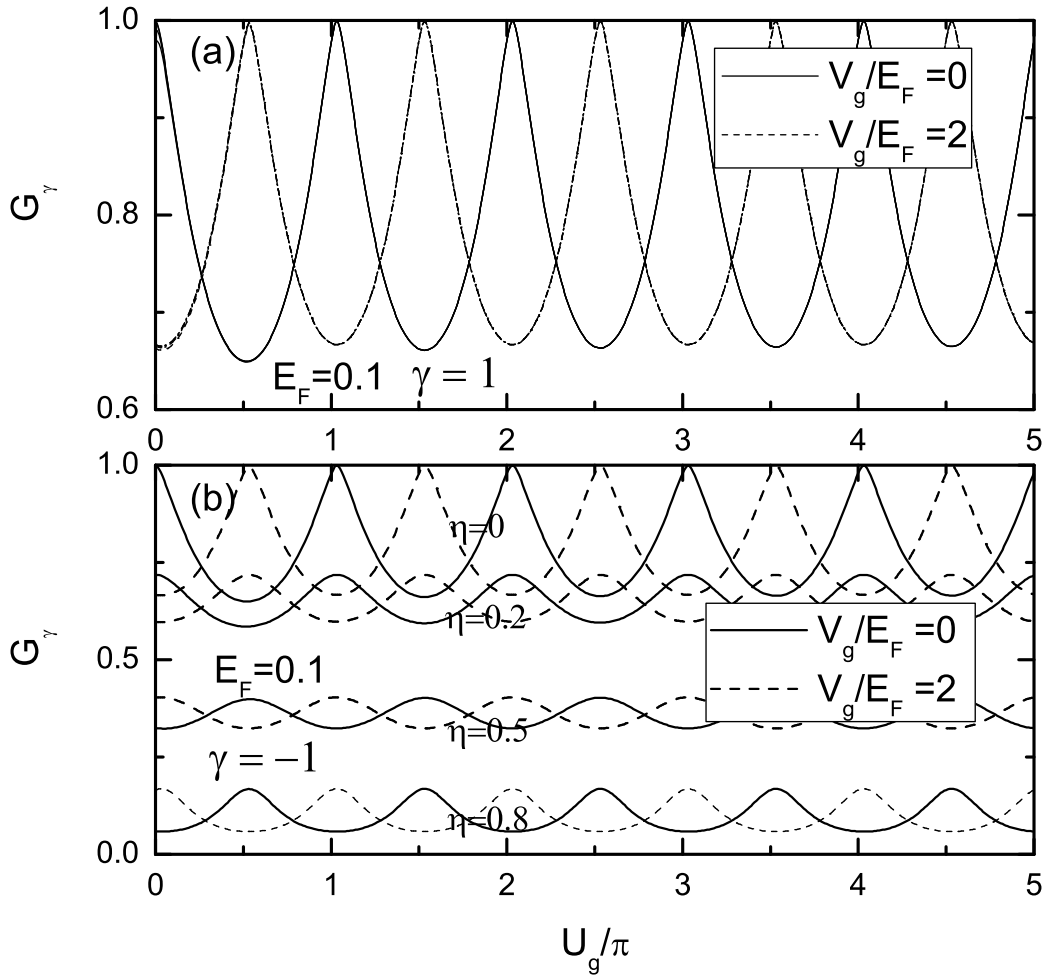}
}
\caption{ Gate voltage dependence of the conductances with a
\textbf{ P} ($\gamma=1$) and  \textbf{AP} ($\gamma=-1$)
configuration for four different  values $\eta=0, 0.2, 0.5$, and
$0.8$. The solid lines are for $V_g /E_F =0$ while the dashed lines
are for $V_g /E_F =2$. The value of the other parameter is $E_F
=0.1$.  }\label{FIG:2eps}       
\end{figure}

In order to observe the effect of the exchange field $\eta$ on the
conductance, in Fig.3 we show the gate voltage dependence of the
conductances with a \textbf{ P} ($\gamma=1$) and \textbf{AP}
($\gamma=-1$) configuration for four different  values $\eta=0, 0.2,
0.5$, and $0.8$. The solid lines are for $V_g /E_F =0$ while the
dashed lines are for $V_g /E_F =2$. The value of the other parameter
is $E_F =0.1$. A similar tendency to Fig. 2 is seen in Fig. 3. It is
easily seen that the exchange field $\eta$ weakly affects the
conductance of carriers for $\gamma=1$ but profoundly influences the
conductance of carriers for $\gamma=-1$.  For $\gamma=-1$,
$G_\gamma$ is suppressed obviously by increasing the value $\eta$.
Due to current conservation, the factor $\Re(\cos \theta_{F_2}/\cos
\theta_{F_1})$ must be real and then we have $\sin \theta_{F_1}=\pm
\sin \theta_{F_2}+ 2 \eta$ where sign + (or -) corresponds to $V_g
/E_F =0$ (or 2). We can see $2 \eta-1\leq \sin \theta_{F_1} \leq 1$
and $2 \eta-1\leq \sin (\mp\theta_{F_2}) \leq 1$ where sign - (or +)
corresponds to $V_g /E_F =0$ (or 2). Thus we find the ranges of the
angle-allowable $\theta_{F_1}$ and $\theta_{F_2}$ depend on $\eta$.
The transmission is nonzero only for $\theta_{F_1}$ and
$\theta_{F_2}$ in these ranges and vanishes for $\eta \geq 1$. The
number of channels decreases with increasing of $\eta$, so we can
see that $G_\gamma$ dramatically decreases with the increase of
$\eta$ for $\gamma=-1$. Noting that the $\eta\geq 1$ for
$\gamma=-1$, the conductance of carriers is forbidden, which implies
anomalous magnetoresistance in this topological spin valve.
\begin{figure}
\resizebox{1.0\columnwidth}{!}{%
  \includegraphics{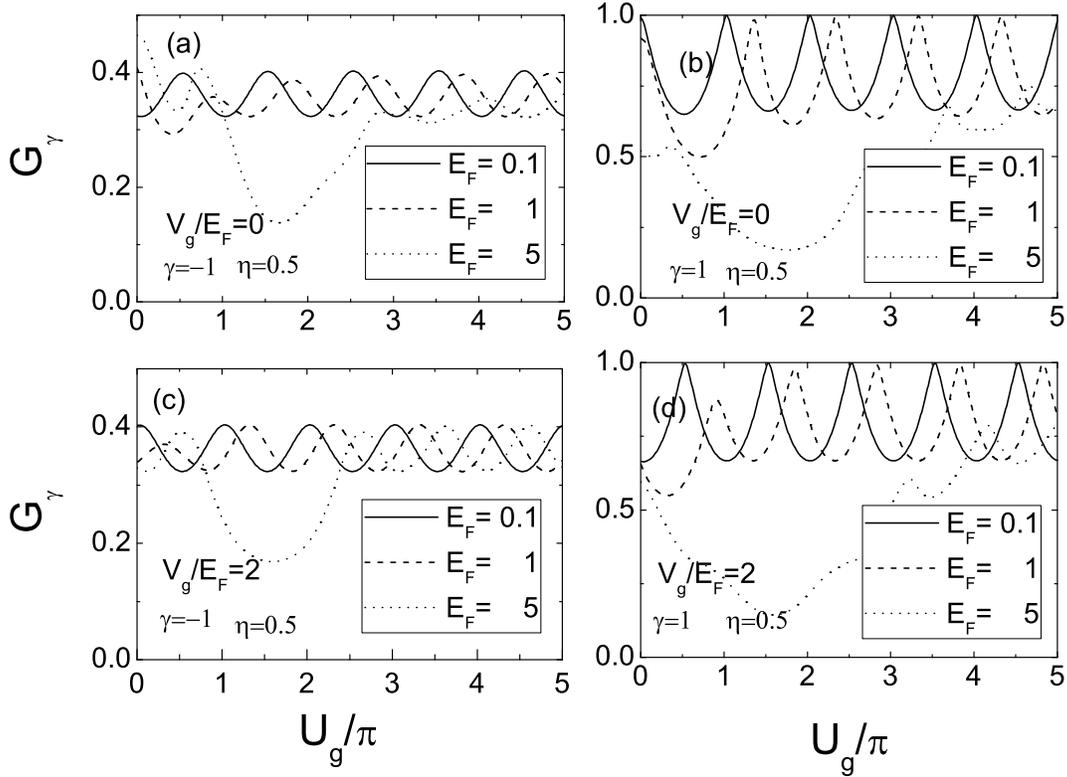}
}
\caption{ Gate voltage dependence of the conductances with a
\textbf{ P} ($\gamma=1$) and  \textbf{AP} ($\gamma=-1$)
configuration for three different  values $E_F = 0.1, 1.0$, and
$5.0$. In (a) and (b), the $V_g$ is set as $V_g /E_F =0$ while in
(c) and (d) the $V_g$ is set as $V_g /E_F =2$. The value of the
other parameter is $\eta =0.5$. }\label{FIG:2eps}       
\end{figure}

In Fig. 4, we show the gate voltage dependence of the conductances
with a \textbf{ P} ($\gamma=1$) and  \textbf{AP} ($\gamma=-1$)
configuration for three different  values $E_F = 0.1, 1.0$, and
$5.0$. In (a) and (b), the $V_g$ is set as $V_g /E_F =0$ while in
(c) and (d) the $V_g$ is set as $V_g /E_F =2$. The value of the
other parameter is $\eta =0.5$. For $E_F =0.1$, we can see that the
$\pi$ periodicity appears. However, the $\pi$ periodicity is broken
for $E_F =1$ (or $5$) because the condition $U_g \gg E_F$ is not
satisfied for the smaller $U_g$.  Nevertheless, we get the $\pi$
periodicity of conductance again by choosing a bigger $U_g$ for the
bigger $E_F$. Furthermore, we find that the minimum of the
conductance will appear when the gate voltage arrives at a certain
value. It is easily seen that the minimum of the conductance shifts
to the right with increasing of the Fermi energy.  The larger the
Fermi energy is, the smaller the minimum of the conductance is. This
phenomena is very obvious for the  \textbf{ P} ($\gamma=1$)
configuration [see in figs.4 (a) and (d)]. From Figs.4 and 5, we
find that the conductance at the parallel (P) configuration can be
much bigger than that at the antiparallel (AP) configuration.
However it may be not satisfied for the larger Fermi energy when the
gate voltage is not bigger enough. We find that there is a
Fabry-Perot-like electron interference in the \textbf{F/N/F} tunnel
junction, which is deposited on the top of a topological surface.
The two ferromagnetic electrodes and the barrier can compose a
Fabry-Perot resonator ${[33,34]}$. The transmitted electron waves in
this resonator can be reflected by the two ferromagnetic electrodes.
The electron waves undergo multiple reflections back and forth along
the resonator between the two ferromagnetic electrodes. The
conductance oscillations are caused by the interference of electron
waves among the modes of the channel-allowable. When the gate
voltage $U_g$ is larger than the Fermi energy $E_F$, the round trip
between the two ferromagnetic electrodes adds a further phase change
$\delta\sim 4\pi/\lambda $ where the Fermi wavelength $\lambda\sim
2\pi/U_g$ because of the value $\theta\sim 0$. When the round trip
between the two ferromagnetic electrodes is equal to the a multiple
of wavelength, the quantum interference happens. This implies that
the oscillation period is equal to $\triangle U_g =\pi$.
\begin{figure}
\resizebox{1.0\columnwidth}{!}{%
  \includegraphics{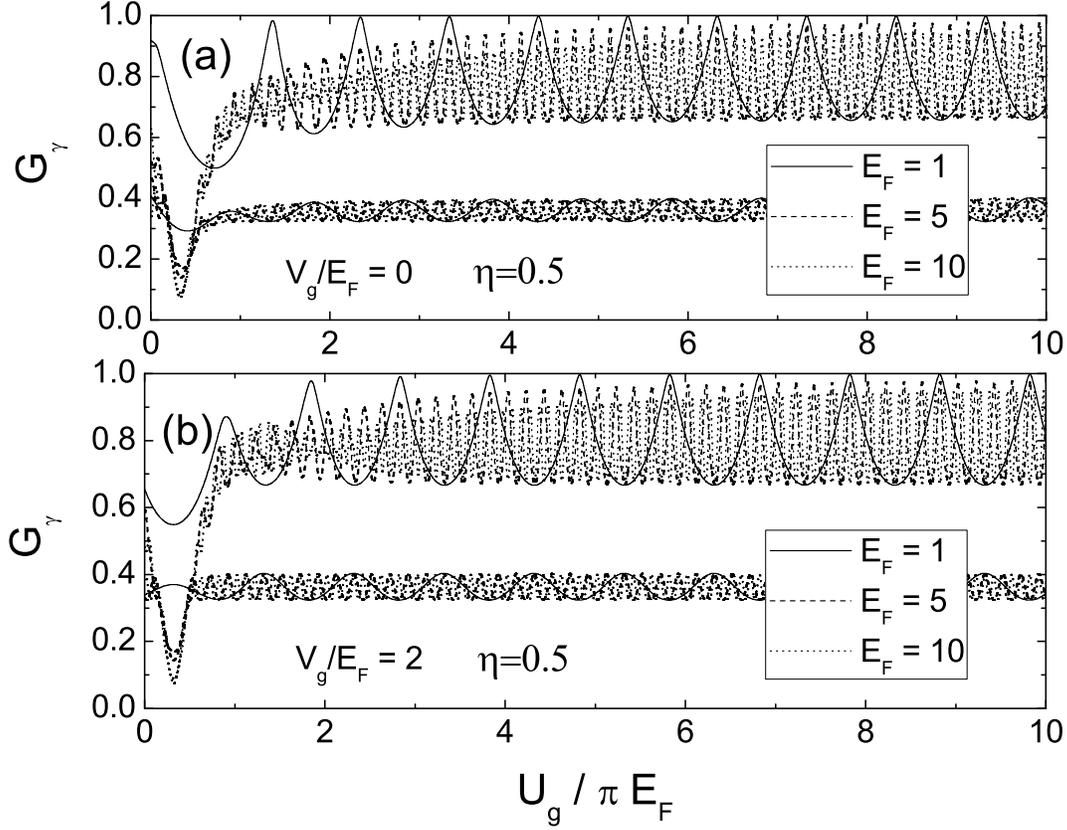}
}
\caption{ Gate voltage dependence of the conductances with a
\textbf{ P} ($\gamma=1$) and  \textbf{AP} ($\gamma=-1$)
configuration for three different  values $E_F = 1.0, 5.0$, and
$10.0$. In (a) , the $V_g$ is set as $V_g /E_F =0$ while in (b)
$V_g$ is set as $V_g /E_F =2$. The value of the other parameter is
$\eta =0.5$.  }\label{FIG:2eps}       
\end{figure}

In order to investigate the solution of the standard electron
described by the Schr$\ddot{o}$dinger equation with a parabolic band
structure, we consider a \textbf{2DEG} in (x,y) plane with a
magnetic field B in the z direction as described in Refs.[26-28].
Thus we can fabricate a F/N/F tunneling based on the \textbf{2DEG}.
We can apply all the relevant quantities in dimensionless units,
which are the same with the Ref. 28. So we can define the value
$\Delta = M_0= B$ where $M_0$ is the the effective exchange field
corresponding to a \textbf{F}/\textbf{N}/\textbf{F} tunneling based
on the\textbf{ TI} and $B$ is the magnetic field corresponding to a
\textbf{F}/\textbf{N}/\textbf{F} tunneling based on \textbf{ 2DEG}.
Nevertheless, we ignore the splitting of energy induced by the spin
of electron. As described in Fig.1, we set the left electrode
potential $V_1=0$. 
For the electron with parabolic spectrum,  ${E_F}  \sin{\theta
_{{F_1}}}$  in Eq.(6) should be replaced by
$\sqrt{2{E_F}}\sin{\theta _{{F_1}}}$. Then the continuity of the
wave function
 gives the transmission coefficient
 \begin{eqnarray*}
 {t_\gamma } = \frac{{2{\kern 1pt} {k_1}{\kern 1pt} {q_x}}}{{ - {q_x}{\kern 1pt}
 ({k_1} + {k_3})\cos {\kern 1pt} {\kern 1pt} ({q_x}{\kern 1pt} L){\kern 1pt}  +
 i{\kern 1pt} {\kern 1pt} ({k_1}{k_3} + q_x^2)s{\kern 1pt} in{\kern 1pt}
 ({q_x}{\kern 1pt} L){\kern 1pt} {\kern 1pt}
 }},
\end{eqnarray*}
and transmission probability ${T_\gamma } = {\left| {{t_\gamma }}
\right|^2}{\kern 1pt} {\kern 1pt} {\mathop{\rm \Re}\nolimits}
({k_3}/{k_1})$ where ${k_1} = \sqrt {2{\kern 1pt} {\kern 1pt}
({E_F}{\kern 1pt}  - {V_1}){\kern 1pt} {\kern 1pt} } {\kern 1pt}
{\kern 1pt} {\kern 1pt} {\kern 1pt} \cos {\kern 1pt} {\kern 1pt}
{\kern 1pt} {\theta _{{F_1}}}$, ${q_x} = \sqrt {2{\kern 1pt} {\kern
1pt} ({E_F}{\kern 1pt}  - {V_2}){\kern 1pt} {\kern 1pt} } \cos
{\kern 1pt} {\kern 1pt} {\kern 1pt} \theta $ and ${k_3} = \sqrt
{2{\kern 1pt} {\kern 1pt} ({E_F}{\kern 1pt}  - {V_3}){\kern 1pt}
{\kern 1pt} } {\kern 1pt} {\kern 1pt} {\kern 1pt} \cos {\kern 1pt}
{\kern 1pt} {\theta _{{F_3}}}$. We can easily see that ${T_\gamma }
\equiv 0$ when the Fermi energy is smaller than the right electrode
voltage ${V_3}$ , which is different from the case of Dirac band
structure. In Fig.6, we show the wave vector $k_y$ dependence of the
transmission probability with a \textbf{ P} ($\gamma=1$) and
\textbf{AP} ($\gamma=-1$) configuration for Dirac electrons shown in
(a) and the standard electrons shown in (b). The values of the other
parameter are $\Delta=0.5$,  $V_2= U_g=0$, $V_3=V_g$ and
 $k_x = k_{x_1} = k_1 = 2$.    We can find that transmission is
significantly more pronounced for Dirac electrons than for the usual
electrons. Compared with the  \textbf{ P} ($\gamma=1$) configuration
, it can be seen that the channel of electron transporting from the
left electrode to the right electron is suppressed especially for
the standard electron. It is easily seen that the variety of the
tunneling conductance of F/N/F with the change of the gate voltage
is obviously different between the Dirac electron and standard
electron [see in Fig.7]. For the standard electron, the conductance
will decrease monotonously with the increase of the gate voltage
because of evanescent wave modes. However, we can see that the
conductance with a $\pi$ periodicity appears as the gate voltage is
large enough for the Dirac electron. On the one hand,  Dirac
confined electron exhibits  a jittering motion called
¡°Zitterbewegung¡±, originating from the interference of states with
positive and negative energy. On the other hand, the transmitted
Dirac electron waves in this resonator can be reflected by the two
ferromagnetic electrodes one after another, so the phase
interference will appear. As a result, this implies the existence of
Fabry-Perot-like electron interference in a \textbf{F/N/F} tunneling
based on the \textbf{TI}.

\begin{figure}
\resizebox{1.1\columnwidth}{!}{%
  \includegraphics{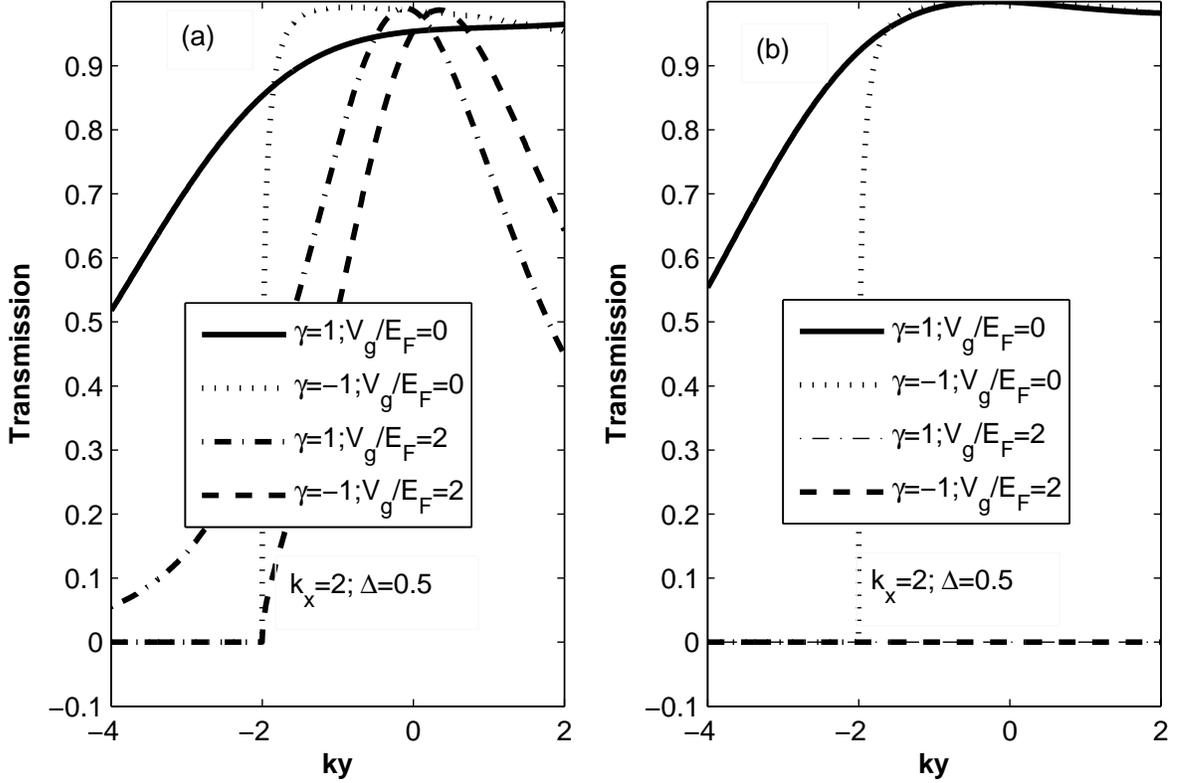}
 }
\caption{ Wave vector $k_y$ dependence of the transmission
probability with a \textbf{ P} ($\gamma=1$) and  \textbf{AP}
($\gamma=-1$) configuration for the Dirac electrons shown in (a) and
the standard electrons shown in (b). The values of the other
parameter are $\Delta=0.5$,  $V_2= U_g=0$, $V_3=V_g$ and
 $k_x = k_{x_1} = k_1 = 2$.  }\label{FIG:2eps}       
\end{figure}

\begin{figure}
\resizebox{1.0\columnwidth}{!}{%
  \includegraphics{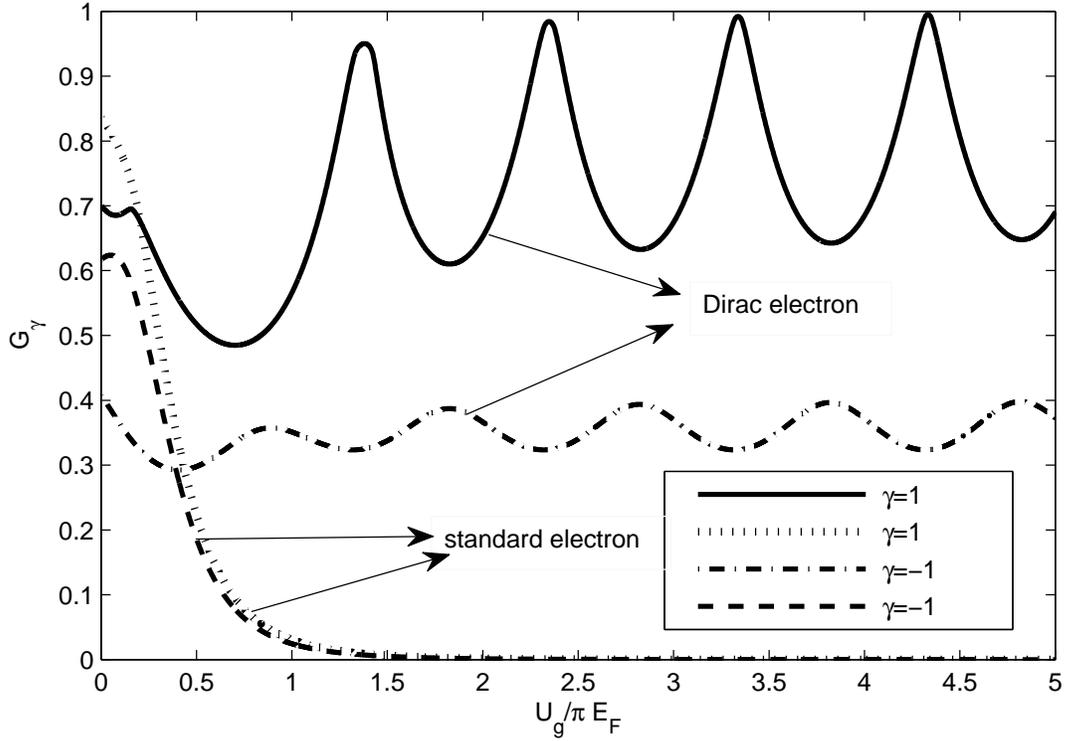}
 }
\caption{ Gate voltage $U_g$ dependence of the conductance with a
\textbf{ P} ($\gamma=1$) and  \textbf{AP} ($\gamma=-1$)
configuration for the Dirac electrons  and the standard electrons.
The values of the other parameter are $\eta=0.5$, $V_3=V_g=0$ and
 $E_F =1.0$.  }\label{FIG:2eps}       
\end{figure}

\section{Conclusion}
In summary, we have theoretically investigated transport features of
Dirac electrons on the surface of a three-dimensional \textbf{TI}
under the modulation of a exchange field provided by an \textbf{FI}
stripes. We find that the conductance at the  \textbf{P}
configuration can be much bigger than that at the \textbf{AP}
configuration.  Compared \textbf{P } with \textbf{ AP}
configuration,  there exists a shift of phase which can be tuned by
gate voltage. We find that the exchange field weakly affects the
conductance of carriers for \textbf{P} configuration but can
dramatically suppress the conductance of carriers for \textbf{AP}
configuration. This controllable electron transport  implies
anomalous magnetoresistance in this topological spin valve, which
may contribute to the development of spintronics .

\vskip0.5cm

\textbf{Acknowledgments}: This work was supported by the National
Natural Science Foundation of China under Grants No. 10174024 and
No. 10474025.

\end{document}